\documentclass[10pt,english,a4paper,nofootinbib,twocolumn]{revtex4}
\usepackage[normalem]{ulem}

\usepackage{amsmath,amssymb,amsfonts}
\usepackage{enumitem}
\usepackage{epsfig}
\usepackage{latexsym}
\usepackage{xcolor}
\usepackage[colorlinks=true,citecolor=blue]{hyperref}
\usepackage[displaymath, tightpage]{preview}
\usepackage{comment}

\usepackage{subfig}
\makeatletter

\@ifundefined{textcolor}{}
{%
 \definecolor{BLACK}{gray}{0}
 \definecolor{WHITE}{gray}{1}
 \definecolor{RED}{rgb}{1,0,0}
 \definecolor{GREEN}{rgb}{0,1,0}
 \definecolor{BLUE}{rgb}{0,0,1}
 \definecolor{CYAN}{cmyk}{1,0,0,0}
 \definecolor{MAGENTA}{cmyk}{0,1,0,0}
 \definecolor{YELLOW}{cmyk}{0,0,1,0}
 \definecolor{PURPLE}{rgb}{0.5,0,0.5}
 }

\newcommand{\bite}{\begin{itemize}}
\newcommand{\eat}{\end{itemize}}
\newcommand{\beq}{\begin{equation}}
\newcommand{\eeq}{\end{equation}}

\newcommand{\beqa}{\begin{align}}
\newcommand{\eeqa}{\end{align}}
\newcommand{\barr}{\begin{array}}
\newcommand{\earr}{\end{array}}

\newcommand{\ie}{\emph{i.e.}}

\newcommand{\M}{\mathcal{M}}

\newcommand{\mb}[1]{\mathbf{#1}}
\newcommand{\mc}[1]{\mathcal{#1}}
\newcommand{\mbb}[1]{\mathbb{#1}}
\newcommand{\mf}[1]{\mathfrak{#1}}

\newcommand{\vect}[1]{\boldsymbol{#1}}
\newcommand{\expect}[1]{\langle #1\rangle}

\newcommand{\supersc}[1]{$^{\textrm{#1}}$}

\makeatother

\begin{document}

\renewcommand\makeindex

\title{Non-abelian Gauge Fields from Defects in Spin-Networks}
\author{Deepak Vaid}
\email{dvaid79@gmail.com}
\affiliation{National Institute of Technology Karnataka (NITK), Surathkal, Karnataka, India}

\date{\today}

\begin{abstract}
\emph{Effective} gauge fields arise in the description of the dynamics of defects in lattices of graphene in condensed matter. The interactions between neighboring nodes of a lattice/spin-network are described by the Hubbard model whose effective field theory at long distances is given by the Dirac equation for an \emph{emergent} gauge field. The spin-networks in question can be used to describe the geometry experienced by a non-inertial observer in flat spacetime moving at a constant acceleration in a given direction. We expect such spin-networks to describe the structure of quantum horizons of black holes in loop quantum gravity. We argue that the abelian and non-abelian gauge fields of the Standard Model can be identified with the emergent degrees of freedom required to describe the dynamics of defects in symmetry reduced spin-networks.
\end{abstract}

\maketitle

\tableofcontents


\setlength{\parskip}{0.16cm}
\setlength{\parindent}{0.0cm}

\section{Introduction}\label{sec:intro}

The debate in Quantum Gravity rages around the question of Lorentz invariance and its manifestations or lack thereof in the various approaches towards unification such as string theory and loop quantum gravity. In string theory the problem is that of recovering general diffeomorphism invariance from a quantum theory for extended objects embedded in a given classical background geometry such as Minkowski, deSitter or anti-deSitter spacetime. From the quantum geometry perspective favored by LQG, the very notion of Lorentz invariance becomes suspect at scales at which quantum geometric effects start to become important. Below the scale at which the discreteness of geometry observables such as area and volume becomes significant, it is not clear how one can define the action of Lorentz transformations. The Lorentz group describes the symmetry of a continuum geometry. In a discrete setting such as that of the spin-networks of LQG the Lorentz group should be superseded by a discrete group. Then the challenge would be to show how under a suitable coarse-graining, the discrete symmetries will approach those of the continuous Lorentz group.

Of equal or perhaps greater importance is the challenge of understanding how one can embed the Standard Model of particle physics in a discrete spin-network background. In LQG we associate spins living in representations of $\mf{su}(2)$ with the edges of a graph and intertwiners with its nodes, which allow us to sum all the edges coming into a given node. In theory there is nothing that prevents us from repeating this construction for any arbitrary gauge group. An obvious thought, then, is to extend the spin-network gauge group from $SU(2)$ to $SU(3) \times SU(2)\times U(1)$, so that the ``spins'' can now be identified with excitations of the Standard Model. The problem with this approach is that it takes as fundamental the gauge structure of the Standard Model. Moreoever it banishes the possibility of constructing a picture where elementary particles arise not as representations of some complicated gauge group but as topological structures from a simple underlying graph based substrate.

In this work we argue for a bottom-up approach where the gauge fields and particles of the Standard Model are not \emph{built into} the model, \emph{a priori}, but arise when one considers the effective field theory of the many body degrees of freedom that a graph possesses. In order to simplify the setting we call upon the behavior of accelerated observers in general relativity which allows us to limit our considerations to two-dimensional (planar) graphs rather than graphs of arbitrary topology. Similar considerations have recently been presented by Banks and Fischler \cite{Banks2013Holographic}.

The outline of this work is as follows. In Section \ref{sec:symmetry_reduced} we see how symmetry-reduced spin-networks serve as the basis for describing the spacetime experienced by an accelerated observer in an otherwise isotropic, homogenous, flat background. In Section \ref{sec:emergent} we recall how the effective field theory description, of the dynamics of a 2DEG (two-dimensional electron gas) living on a graphene lattice, is that of massless Dirac fermions in two-dimensions. In Section \ref{sec:defects} we see how the presence of defects necessitates the introduction of dynamic gauge fields which can be identified with the gauge fields of the standard model.

\section{Symmetry Reduced Spin Networks \& Accelerated Observers}\label{sec:symmetry_reduced}

Spin-networks are the natural arena for the microscopic dynamics of quantum gravity, and provide us with the lattice structure on which we can study models such as the Hubbard model and its cousins. Indeed several authors have taken concrete steps in that direction. The general approach of studying the dynamics of various condensed matter systems defined on static and dynamics graphs is referred to as \emph{quantum graphity} \cite{Ansari2008A-statistical,Caravelli2011Trapped,Chen2012Statistical,Hamma2010A-quantum,Konopka2008Quantum,Konopka2008Statistical,Quach2012Domain}. In general, one can define a spin-network state on \emph{any} graph and it is unclear how one can bring this arbitrariness under control, though see \cite{Caravelli2011Trapped,Chen2012Statistical,Hamma2010A-quantum} for attempts at tackling this question. One solution lies in considering \emph{symmetry reduced} graphs, \emph{i.e.}, those which satisfy foliation invariance, and/or approximate rotational and translational invariance as discussed in \cite{Caravelli2011Trapped} which would allow us to construct more complex graphs by taking suitable sums of simpler sub-graphs. 

The best known example of this can be found in the field of Loop Quantum Cosmology (LQC), where the methods of LQG are applied to understanding questions such as the behavior of cosmological spacetimes in the vicinity of singularities such as the ``big bang'' and the ``big crunch''. On the very largest scales the Universe is isotropic and homogenous - this notion is supported by measurements of the CMB spectrum\cite{Hinshaw2012Nine-Year,Collaboration2013Planck}, by measurements of baryon density oscillations, by galaxy surveys which determine the large scale distribution of matter and also by numerical simulation of cosmological structure formation. The spin-network states, which are used to construct the quantum versions of the Hamiltonian and Gauss constraints for a cosmological spacetime, are the so-called symmetry reduced states first studied in this context by Bojowald\cite{Bojowald1999Loop}.

In Section \ref{sec:intro} we will discuss the emergence of a $2+1$ dimensional spacetime containing a gas of non-interacting Dirac fermions as the low-energy effective excitations of the system whose microscopic dynamics consists of ``electrons'' hopping between vertices of a honeycomb lattice. The Hamiltonian is that of the Hubbard model on the honeycomb lattice at ``half-filling'' - \emph{i.e.}, with one electron at each vertex. The question then is, \emph{what role (if any) can this model play in understanding the emergence of a \emph{$3+1$} dimensional spacetime}? The existence of phenomena associated with non-inertial observers in otherwise flat spacetimes can allow us to tackle this question.

Let us consider an observer moving with a constant acceleration in a particular direction which we take to be the $\hat z$-axis for convenience. An observer at rest experiences a local spatial geometry which obeys an $SO(3)$ symmetry (at macroscopic scales). The spacetime for an accelerating observer develops a horizon which breaks this symmetry down to the $SO(2)\times Z_2 $, where the $SO(2)$ corresponds to the observers freedom to rotate in a plane normal to the $\hat z$-axis, and $Z_2$ corresponds to her freedom to move along the positive or negative $\hat z$-axis, which can also be seen as a discrete reflection symmetry across the horizon. That the geometry of accelerated observers has a preferred axis and exhibits cylindrical symmetry can also be seen from the equivalence principle, according to which such an observer experiences the same physics as an observer in a uniform gravitational field whose direction defines the notion of \emph{upwards} and \emph{downwards}. The preferred axis in such a setting is determined by the direction in which freely-falling bodies appear to move in the frame of an observer ``at rest''.

Thus, in order to model the quantum geometry experienced by an accelerated observer, in an otherwise flat spacetime, we need a spin-network which can be seen to possess the symmetries of a spacetime experienced by an accelerated observer - \emph{i.e.}, $SO(2)\times Z_2 $. Such a spin-network is easily furnished by the graph corresponding to the graphite lattice - which consists of graphene sheets stacked in layers on top of each other. Such a graph has a natural symmetry axis normal to the planes containing graphene sheets and parallel to the direction of the stacking. \textbf{[illustration]}.

Now, \emph{a priori}, we only need a graph with a ``cylindrical'' symmetry. Any two-dimensional graphs - with structure different from that of the honeycomb lattice of graphene - stacked on top of each other would satisfy this requirement. However, there are several reasons to prefer the honeycomb lattice. One reason is \emph{convenience}. The analysis of the Hubbard model on the honeycomb lattice is a well-understood problem and one can readily adopt results obtained in the study of graphene to our spin-network setting. A second reason is that of \emph{universality}. Of all the possible two-dimensional (planar) graphs, with approximate rotational invariance, the honeycomb lattice is singled out by a dynamical argument \cite{Atiyah2001The-Geometry,Atiyah2003Polyhedra}

\section{Emergent Lorentz Invariance in Graphene}\label{sec:emergent}

Regardless of the true microscopic symmetries of quantum geometry, what is relevant for our purposes is to understand the ways in which Lorentz invariance can emerge as a symmetry of the \emph{effective} theory describing the low-energy, long-range excitations of quantum geometry. The simplest such models are encountered in the field of condensed matter. An example is the case of graphene whose structure is given by a hexagonal or honeycomb lattice.

Carbon atoms with sp\supersc{3} hybridized orbitals form a honeycomb lattice where each carbon atom is covalently bonded to three neighboring atoms taking up three of the four available orbitals. The fourth orbital is occupied by the remaining carbon valence electron. Thus each site in the lattice has a single electron - for a perfect lattice - with room for one more (each orbital can accommodate two electrons). Graphene is thus said to be at half-filling, because the conduction band is half-filled, making graphene an insulator or semiconductor at best.

The dynamics of the free electron fluid of graphene can be understood by using to a first approximation the tight-binding or Hubbard model Hamiltonian for interaction between each carbon and its three nearest neighbors \cite{Gonzalez1992The-Electronic}. When the continuum limit of the resulting eigenvalue equation is taken one finds that the quasiparticles obey the Dirac equation for massless, spin 1/2 fermions.

The microscopic theory governing the inter-particle interactions is given by the tight-binding Hamiltonian describing the interactions between fermions living on neighboring sites \cite{Gonzalez1992The-Electronic, Neto2008Electronic}:
\begin{align}\label{eqn:hubbard-model}
	H = -t \sum\limits_{\langle i,j \rangle, \sigma} & \left( a_{i,\sigma}^{\dagger} b_{j,\sigma} + h.c.\right) \nonumber \\ 
	& - t' \sum\limits_{\langle\langle i,j \rangle \rangle, \sigma} \left( a^{\dagger}_{i,\sigma} a_{j,\sigma} + b^{\dagger}_{i,\sigma} b_{j,\sigma} + h.c. \right)
\end{align}
where the brackets $ \langle \ldots \rangle $ imply that the sum is over the nearest-neighbors and $ \langle\langle \ldots \rangle \rangle $ imply that the sum is over the next-nearest neighbors. The graphene lattice is made of two interpenetrating triangular lattices \textbf{[illustration]} labeled A and B. The ladder-operators $\{ a^{\dagger}_{i,\sigma}, a_{j,\sigma} \} $ and $ \{ b^{\dagger}_{i,\sigma}, b_{j,\sigma} \} $  are the creation/annihilation operators for sites on sublattices A and B respectively. $ t $ and $ t' $ measure the strength of the nearest-neighbor and next-nearest-neighbor coupling respectively.

The band-structure for the Hamiltonian \ref{eqn:hubbard-model} is of the form \cite{Wallace1947The-Band,Neto2008Electronic}:
\begin{equation}\label{eqn:band-structure}
	E_{\pm}(\vect{k}) = \pm t \sqrt{3 + f(\vect{k})} - t' f(\vect{k})
\end{equation}
Here $ \vect{k} = (k_x, k_y) $ is the wave-vector denoting points in the momentum space and the function $ f(\vect{k}) $ is given by:
\begin{equation}\label{eqn:band-structure-2}
	f(\vect{k}) = 2 \cos \left(\sqrt{3} k_y a\right) + 4 \cos\left(\frac{\sqrt{3}}{2} k_y a\right) \cos\left(\frac{3}{2} k_x a\right)
\end{equation}
where $ a $ is the spacing between nearest-neighbors on the hexagonal lattice. The Brillouin zon (BZ) of the honeycomb lattice possess two points labeled $\vect{K}$ and $\vect{K'}$, at which the band energy vanishes $E_{\pm}(\vect{K}) = E_{\pm}(\vect{K}') = 0$. Thus, after turning off the next-to-nearest neighbor interaction by setting $t'=0$, we can expand \ref{eqn:band-structure} around either of these two points by writing the momentum as $\vect{k}=\vect{K}+\vect{q}$, where $|\vect{q}| \ll |\vect{K}|$. Doing so, we obtain:
\begin{equation}\label{eqn:dirac-point}
	E_{\pm}(\vect{q}) \simeq \pm v_F |\vect{q}| + \mc{O}([q/K]^2)
\end{equation}
where $v_F = 3ta/2$ is called the Fermi velocity. This shows that in the vicinity of the so-called ``Dirac points'', $\vect{K}, \vect{K'}$, the energy dispersion is that of massless fermions whose dynamics would therefore be described by the massless Dirac equation:
\begin{equation}\label{eqn:dirac-eqn}
	\gamma^\mu \partial_\mu \psi = 0
\end{equation}
in $2+1$ dimensions. Consequently excitations, near the ground state of the many-body electron system on the honeycomb lattice behave as, \emph{for all practical purposes}, particles which live in a flat $2+1$ dimensional spacetime and respect the (approximate) Lorentz symmetry of that effective background geometry. Turning on the next-to-nearest neighbor interaction ($t'\neq 0$), shifts the spectrum given in \ref{eqn:dirac-point} by a constant amount, but does not modify the its linear form:
\begin{equation}\label{eqn:modified-dirac-point}
	E_{\pm}(\vect{q}) \simeq 3 t' \pm v_F |\vect{q}| + \mc{O}([q/K]^2)
\end{equation}
Our quasiparticles are now described by the Dirac equation for massive fermions:
\begin{equation}\label{eqn:massive-dirac-eqn}
	(\gamma^\mu \partial_\mu + m) \psi = 0
\end{equation}
where $m \sim 3 t'$. Thus turning on the next-to-nearest neighbor interaction breaks the electron-hole symmetry near the Dirac points and generates a non-zero rest mass for the quasiparticles.

In this manner, we see how the an (approximate) Lorentz symmetry emerges in a discrete many-body system in which such a symmetry was not present to begin with.

\section{Role of defects}\label{sec:defects}

In Section \ref{sec:symmetry_reduced} we discussed how accelerating observers experience a spacetime geometry whose microscopic structure can be identified with that of a honeycomb lattice. In Section \ref{sec:emergent} we saw how the quasiparticle excitations of particles hopping between various sites of this lattice can be identified with massless (or massive if $t'\neq 0$) Dirac fermions in the low-energy limit. In Nature, one does not expect to encounter perfect lattices. In fact, disorder in the form of defects are \emph{essential} for phenomena such as the quantization of hall plateaus observed in the limits of very strong magnetic fields and or very high carrier concentrations \cite{Jain1989Compositefermion}. One would expect that the honeycomb lattices which describe the quantum geometry of a horizon would also have defects. In fact, the presence of defects is \emph{required} if the horizons have any curvature. A perfect honeycomb lattice can only define a flat manifold. In order for the manifold to have positive (negative) curvature, the lattice must have defects where hexagons are replaced by pentagons (septagons).

The dual of the honeycomb lattice is a triangular lattice $\mc{L}^*$ which serves as a triangulation of the horizon. Curvature is concentrated at the vertices of $\mc{L}^*$ and is measured by the deficit angle \textbf{[illustration]} around each vertex. In the defect-free case, six equilateral triangles meet at each vertex of $\mc{L}^*$. The sum of the angles subtended by each of the six triangles at the vertex is $ \sum_{i(v)} \theta_i = \pi$. Removing a triangle reduces this sum $\pi - \delta \theta$ and adding a triangle increases this sum by some amount $\pi + \delta \theta$ \textbf{[illustrate]}. $\delta \theta$ is referred to as the \emph{angular deficit} or \emph{deficit angle} around the given vertex. If $\delta \theta > 0$ ($\delta \theta < 0$), the curvature at the vertex is positive (resp. negative). Creating a defect with positive (negative) curvature at a vertex $ v \in \mc{L}^*$ corresponds to replacing the hexagon dual to $v$ in the honeycomb lattice $\mc{L}$ by a pentagon (resp. heptagon).

The simplest defects will correspond to replacing hexagons by pentagons or septagons - i.e. destroying or creating an edge in the graphene lattice. Since each edge is shared by two hexagons, destroying or creating an edge leads to the creation of \emph{pairs} of pentagons or septagons.\footnote{There is another way to create a defect. Cut the graphene lattice starting at some vertex $ v_0 $ and out to infinity. Then one can either add or remove a slice of hexagons to create a conical singularity centered at $ v_0 $ with either positive or negative curvature respectively.}

The presence of defects and disorder in the underlying lattice induces an effective interaction between these quasiparticles. This interaction can be modeled by the introduction of a gauge field $A_\mu$, as shown in \cite{Gonzalez1992The-Electronic}, in the Dirac equation description (\ref{eqn:dirac-eqn} and \ref{eqn:massive-dirac-eqn}) of graphene quasiparticles:
\begin{equation}\label{eqn:gauged-dirac}
	(\gamma^\mu \nabla_\mu + m) \psi = 0
\end{equation}
where $\nabla_\mu = \partial_\mu + i g A_\mu$ is the \emph{gauge covariant} derivative operator, and $g$ measures the strength of the gauge field. However there will be interactions not only between pairs of quasiparticles but also between quasiparticle-defect pairs and defect-defect pairs. Interactions between defects will induce terms in the effective action involving the curvature $F_{\mu\nu}$ of the emergent gauge field:
\begin{equation}\label{eqn:effective-action}
	S = \int d^3 x \, \bar{\psi} (\gamma^\mu \nabla_\mu + m) \psi + F^{\mu\nu} F_{\mu\nu}
\end{equation}
Consequently, in addition to the Dirac equation, obtained by varying $S$ w.r.t. $\psi$, we would have another equation which describes the dynamics of $A_\mu$ and is obtained by varying $S$ w.r.t. $A_\mu$:
\begin{equation}\label{eqn:gauge-eom}
	\nabla_\mu F^{\mu\nu} = j^\mu
\end{equation}
where $j^\mu = ig \bar\psi \gamma^\mu \psi$ is the quasiparticle current.

Furthermore, the BZ of the graphene lattice possesses a ``valley symmetry'' which requires the introduction of ``internal'' indices for the gauge field leading us from an abelian to a non-abelian setting $A_\mu \rightarrow A_\mu^i$, where the gauge field now lives in the Lie-algebra of some gauge group ($SU(2)$ in the example discussed in \cite{Gonzalez1992The-Electronic}) and the index $i$ labels the generators of the gauge group.

Thus the presence of imperfections and defects in the lattice describing the quantum state of geometry, automatically leads to the following ingredients:
\begin{enumerate}
	\item \textbf{Quasiparticles} (massless or massive) living in a $2+1$ dimensional spacetime.
	\item \textbf{Lorentz invariance} of the resulting system due to the fact that quasiparticles satisfy the Dirac equation.
	\item \textbf{Gauge fields - both abelian and non-abelian} required to model interactions between quasiparticles and lattice imperfections, and \ldots
	\item \textbf{Gauge bosons} which arise as the propagating degrees of freedom of the gauge field and which mediate interactions between pairs of defects.
\end{enumerate}

In this manner we see how starting from a simple, discrete many-body interacting system defined on a two-dimensional lattice, we obtain all the ingredients required for modelling both particles and the gauge fields which mediate interactions between them. With suitable adjustments it is quite possible that one might even be able to obtain something resembling the structure of the standard model starting from such a basic framework. This would provide us with a concrete avenue to integrate the standard model with the spin-network picture of quantum geometry inherited from the LQG approach.



One can reverse the above reasoning to argue that turning on a magnetic field in a vacuum in a flat spacet-time leads to the creation of defects in the hexagonal spin-network which describes the flat geometry in the absence of a magnetic field. These defects change the ground state of the geometry of this vacuum. It is reasonable to assume that the area density of geometric defects increases with the increasing density of magnetic flux. Secondly, in a lattice all defects have some \emph{finite} size. At small magnetic fields the number of defects is small and their dynamics is that of a dilute weakly interacting gas. However, as we turn up the magnetic field strength the number of defects proliferates and at some critical field strength $B_C$ the defects will sufficient overlap such that they must condense to form a \emph{defect fluid}. Such a picture has also been utilized by Maxim Chernodub \cite{Chernodub2010Superconductivity,Chernodub2011Can-nothing,Chernodub2011Spontaneous} in developing an understanding of the spontaneous formation of a rho-meson condensate in an empty spacetime in the presence of very strong ($10^{16}$ Tesla) magnetic fields.

On a more speculative note, these considerations also open up the possibility that gauge fields could be used to manipulate the geometry of a given region of spacetime. One can imagine a situation where a defect condensate could serve to divide a spacetime region into two or more regions. The two-dimensional surface itself would contain the defect condensate. However, the gauge fields generated by the defect fluid would necessarily permeate into the surrounding three-dimensional neighbourhood. \emph{If} these gauge fields can be identified with those which describe the dynamics of the gravitational fields in the LQG framework, then it follows, that the geometry of the three-dimensional neighbourhood of the surface would necessarily be perturbed by such induced gauge fields. In this way, one can imagine constructing concrete, experimentally realizable systems which exhibit holographic behaviour in the laboratory. Though, it sounds far fetched at the present moment, such systems could \emph{conceivably} be utilized to construct so-called \emph{anti-gravity} devices depicted in the science fiction literature. Whether or not something of this sort is theoretically possible, let alone experimentally feasible, requires far more investigation.

%


\textbf{Author's note:} \emph{Just prior to the posting of the first version of this article on arXiv, two articles \cite{Hossenfelder2013aPhenomenology,Hossenfelder2013bPhenomenology}, investigating the phenomenological consequences of defects in spacetime arising from an underlying non-geometric theory of gravity, were posted on arXiv. Though there is likely to be significant thematic overlap between our work and these papers, our ideas were developed independently and separately from those in the mentioned works. Taken together, the present work and \cite{Hossenfelder2013aPhenomenology,Hossenfelder2013bPhenomenology} provide investigations into the same question from two different perspectives and thus are likely to complement each other.}


\begin{acknowledgements}
I would like to thank my colleagues (AKM, HSN and PPD) and friends (SG and the residents of A-wing, 1st floor, Mega 2 hostel) at NITK for providing a supportive academic environment, which makes it possible to continue working on research, while remaining engaged in, the primary activity of our institution, teaching.
\end{acknowledgements}


\bibliographystyle{kp}

\bibliography{../bib_library}

\begingroup\raggedright\begin{thebibliography}{22}
\expandafter\ifx\csname natexlab\endcsname\relax\def\natexlab#1{#1}\fi

\bibitem[Banks and Fischler(2013)]{Banks2013Holographic}
T.~Banks and W.~Fischler, ``Holographic theory of accelerated observers, the
  s-matrix, and the emergence of effective field theory'', 2013.

\bibitem[Ansari and Markopoulou(2008)]{Ansari2008A-statistical}
M.~H. Ansari and F.~Markopoulou, ``A statistical formalism of causal dynamical
  triangulations'',  \href{http://xxx.lanl.gov/abs/hep-th/0505165}{{\ttfamily
  hep-th/0505165}}.

\bibitem[Caravelli et~al.(2011)Caravelli, Hamma, Markopoulou, and
  Riera]{Caravelli2011Trapped}
F.~Caravelli, A.~Hamma, F.~Markopoulou, and A.~Riera, ``Trapped surfaces and
  emergent curved space in the {Bose-Hubbard} model'', 2011.

\bibitem[Chen and Plotkin(2012)]{Chen2012Statistical}
S.~Chen and S.~S. Plotkin, ``Statistical mechanics of graph models and their
  implications for emergent manifolds'', 2012.

\bibitem[Hamma et~al.(2010)Hamma, Markopoulou, Lloyd, Caravelli, Severini, and
  Markstrom]{Hamma2010A-quantum}
A.~Hamma, F.~Markopoulou, S.~Lloyd, F.~Caravelli, S.~Severini, and
  K.~Markstrom, ``A quantum {Bose-Hubbard} model with evolving graph as toy
  model for emergent spacetime'', {\em Physical Review D} {\bfseries 81}
  (2010), no.~10,  \href{http://xxx.lanl.gov/abs/0911.5075}{{\ttfamily
  0911.5075}}.

\bibitem[Konopka et~al.(2008)Konopka, Markopoulou, and
  Severini]{Konopka2008Quantum}
T.~Konopka, F.~Markopoulou, and S.~Severini, ``Quantum graphity: a model of
  emergent locality'',  \href{http://xxx.lanl.gov/abs/0801.0861}{{\ttfamily
  0801.0861}}.

\bibitem[Konopka(2008)]{Konopka2008Statistical}
T.~Konopka, ``Statistical mechanics of graphity models'',
  \href{http://xxx.lanl.gov/abs/0805.2283}{{\ttfamily 0805.2283}}.

\bibitem[Quach et~al.(2012)Quach, Su, Martin, and Greentree]{Quach2012Domain}
J.~Q. Quach, C.-H. Su, A.~M. Martin, and A.~D. Greentree, ``Domain structures
  in quantum graphity'',  \href{http://xxx.lanl.gov/abs/1203.5367}{{\ttfamily
  1203.5367}}.

\bibitem[Hinshaw et~al.(2012)Hinshaw, Larson, Komatsu, Spergel, Bennett,
  Dunkley, Nolta, Halpern, Hill, Odegard, Page, Smith, Weiland, Gold, Jarosik,
  Kogut, Limon, Meyer, Tucker, Wollack, and Wright]{Hinshaw2012Nine-Year}
G.~Hinshaw, D.~Larson, E.~Komatsu, D.~N. Spergel, C.~L. Bennett, J.~Dunkley,
  M.~R. Nolta, M.~Halpern, R.~S. Hill, N.~Odegard, L.~Page, K.~M. Smith, J.~L.
  Weiland, B.~Gold, N.~Jarosik, A.~Kogut, M.~Limon, S.~S. Meyer, G.~S. Tucker,
  E.~Wollack, and E.~L. Wright, ``{Nine-Year} wilkinson microwave anisotropy
  probe ({WMAP}) observations: Cosmological parameter results'', 2012.

\bibitem[Collaboration et~al.(2013)Collaboration, Ade, Aghanim,
  Armitage-Caplan, Arnaud, Ashdown, Atrio-Barandela, Aumont, Baccigalupi,
  Banday, Barreiro, Bartlett, Bartolo, Battaner, Battye, Benabed, Beno\^{i}t,
  Benoit-L\'{e}vy, Bernard, Bersanelli, Bielewicz, Bobin, Bock, Bonaldi,
  Bonavera, Bond, Borrill, Bouchet, Bridges, Bucher, Burigana, Butler, Cardoso,
  Catalano, Challinor, Chamballu, Chary, Chiang, Chiang, Christensen, Church,
  Clements, Colombi, Colombo, Couchot, Coulais, Crill, Cruz, Curto, Cuttaia,
  Danese, Davies, Davis, de~Bernardis, de~Rosa, de~Zotti, Delabrouille,
  Delouis, D\'{e}sert, Diego, Dole, Donzelli, Dor\'{e}, Douspis, Ducout, Dupac,
  Efstathiou, Elsner, En{\ss}lin, Eriksen, Fantaye, Fergusson, Finelli, Forni,
  Frailis, Franceschi, Frommert, Galeotta, Ganga, Giard, Giardino,
  Giraud-H\'{e}raud, Gonz\'{a}lez-Nuevo, G\'{o}rski, Gratton, Gregorio,
  Gruppuso, Hansen, Hansen, Hanson, Harrison, Helou, Henrot-Versill\'{e},
  Hern\'{a}ndez-Monteagudo, Herranz, Hildebrandt, Hivon, Hobson, Holmes,
  Hornstrup, Hovest, Huffenberger, Jaffe, Jaffe, Jones, Juvela, Keih\"{a}nen,
  Keskitalo, Kim, Kisner, Knoche, Knox, Kunz, Kurki-Suonio, Lagache,
  L\"{a}hteenm\"{a}ki, Lamarre, Lasenby, Laureijs, Lawrence, Leahy, Leonardi,
  Leroy, Lesgourgues, Liguori, Lilje, Linden-V{\o}rnle, L\'{o}pez-Caniego,
  Lubin, Mac\'{\i}as-P\'{e}rez, Maffei, Maino, Mandolesi, Mangilli, Marinucci,
  Maris, Marshall, Martin, Mart\'{\i}nez-Gonz\'{a}lez, Masi, Matarrese,
  Matthai, Mazzotta, McEwen, Meinhold, Melchiorri, Mendes, Mennella,
  Migliaccio, Mikkelsen, Mitra, Miville-Desch\^{e}nes, Molinari, Moneti,
  Montier, Morgante, Mortlock, Moss, Munshi, Naselsky, Nati, Natoli,
  Netterfield, N{\o}rgaard-Nielsen, Noviello, Novikov, Novikov, Osborne,
  Oxborrow, Paci, Pagano, Pajot, Paoletti, Pasian, Patanchon, Peiris,
  Perdereau, Perotto, Perrotta, Piacentini, Piat, Pierpaoli, Pietrobon,
  Plaszczynski, Pointecouteau, Pogosyan, Polenta, Ponthieu, Popa, Poutanen,
  Pratt, Pr\'{e}zeau, Prunet, Puget, Rachen, R\"{a}th, Rebolo, Reinecke,
  Remazeilles, Renault, Renzi, Ricciardi, Riller, Ristorcelli, Rocha, Rosset,
  Rotti, Roudier, Rubi\~{n}o Mart\'{\i}n, Rusholme, Sandri, Santos, Savini,
  Scott, Seiffert, Shellard, Souradeep, Spencer, Starck, Stolyarov, Stompor,
  Sudiwala, Sureau, Sutter, Sutton, Suur-Uski, Sygnet, Tauber, Tavagnacco,
  Terenzi, Toffolatti, Tomasi, Tristram, Tucci, Tuovinen, T\"{u}rler,
  Valenziano, Valiviita, Van~Tent, Varis, Vielva, Villa, Vittorio, Wade,
  Wandelt, Wehus, White, Wilkinson, Yvon, Zacchei, and
  Zonca]{Collaboration2013Planck}
P.~Collaboration, P.~A.~R. Ade, N.~Aghanim, C.~Armitage-Caplan, M.~Arnaud,
  M.~Ashdown, F.~Atrio-Barandela, J.~Aumont, C.~Baccigalupi, A.~J. Banday,
  R.~B. Barreiro, J.~G. Bartlett, N.~Bartolo, E.~Battaner, R.~Battye,
  K.~Benabed, A.~Beno\^{i}t, A.~Benoit-L\'{e}vy, J.~P. Bernard, M.~Bersanelli,
  P.~Bielewicz, J.~Bobin, J.~J. Bock, A.~Bonaldi, L.~Bonavera, J.~R. Bond,
  J.~Borrill, F.~R. Bouchet, M.~Bridges, M.~Bucher, C.~Burigana, R.~C. Butler,
  J.~F. Cardoso, A.~Catalano, A.~Challinor, A.~Chamballu, R.~R. Chary, L.~Y.
  Chiang, H.~C. Chiang, P.~R. Christensen, S.~Church, D.~L. Clements,
  S.~Colombi, L.~P.~L. Colombo, F.~Couchot, A.~Coulais, B.~P. Crill, M.~Cruz,
  A.~Curto, F.~Cuttaia, L.~Danese, R.~D. Davies, R.~J. Davis, P.~de~Bernardis,
  A.~de~Rosa, G.~de~Zotti, J.~Delabrouille, J.~M. Delouis, F.~X. D\'{e}sert,
  J.~M. Diego, H.~Dole, S.~Donzelli, O.~Dor\'{e}, M.~Douspis, A.~Ducout,
  X.~Dupac, G.~Efstathiou, F.~Elsner, T.~A. En{\ss}lin, H.~K. Eriksen,
  Y.~Fantaye, J.~Fergusson, F.~Finelli, O.~Forni, M.~Frailis, E.~Franceschi,
  M.~Frommert, S.~Galeotta, K.~Ganga, M.~Giard, G.~Giardino,
  Y.~Giraud-H\'{e}raud, J.~Gonz\'{a}lez-Nuevo, K.~M. G\'{o}rski, S.~Gratton,
  A.~Gregorio, A.~Gruppuso, M.~Hansen, F.~K. Hansen, D.~Hanson, D.~Harrison,
  G.~Helou, S.~Henrot-Versill\'{e}, C.~Hern\'{a}ndez-Monteagudo, D.~Herranz,
  S.~R. Hildebrandt, E.~Hivon, M.~Hobson, W.~A. Holmes, A.~Hornstrup,
  W.~Hovest, K.~M. Huffenberger, T.~R. Jaffe, A.~H. Jaffe, W.~C. Jones,
  M.~Juvela, E.~Keih\"{a}nen, R.~Keskitalo, J.~Kim, T.~S. Kisner, J.~Knoche,
  L.~Knox, M.~Kunz, H.~Kurki-Suonio, G.~Lagache, A.~L\"{a}hteenm\"{a}ki, J.~M.
  Lamarre, A.~Lasenby, R.~J. Laureijs, C.~R. Lawrence, J.~P. Leahy,
  R.~Leonardi, C.~Leroy, J.~Lesgourgues, M.~Liguori, P.~B. Lilje,
  M.~Linden-V{\o}rnle, M.~L\'{o}pez-Caniego, P.~M. Lubin, J.~F.
  Mac\'{\i}as-P\'{e}rez, B.~Maffei, D.~Maino, N.~Mandolesi, A.~Mangilli,
  D.~Marinucci, M.~Maris, D.~J. Marshall, P.~G. Martin,
  E.~Mart\'{\i}nez-Gonz\'{a}lez, S.~Masi, S.~Matarrese, F.~Matthai,
  P.~Mazzotta, J.~D. McEwen, P.~R. Meinhold, A.~Melchiorri, L.~Mendes,
  A.~Mennella, M.~Migliaccio, K.~Mikkelsen, S.~Mitra, M.~A.
  Miville-Desch\^{e}nes, D.~Molinari, A.~Moneti, L.~Montier, G.~Morgante,
  D.~Mortlock, A.~Moss, D.~Munshi, P.~Naselsky, F.~Nati, P.~Natoli, C.~B.
  Netterfield, H.~U. N{\o}rgaard-Nielsen, F.~Noviello, D.~Novikov, I.~Novikov,
  S.~Osborne, C.~A. Oxborrow, F.~Paci, L.~Pagano, F.~Pajot, D.~Paoletti,
  F.~Pasian, G.~Patanchon, H.~V. Peiris, O.~Perdereau, L.~Perotto, F.~Perrotta,
  F.~Piacentini, M.~Piat, E.~Pierpaoli, D.~Pietrobon, S.~Plaszczynski,
  E.~Pointecouteau, D.~Pogosyan, G.~Polenta, N.~Ponthieu, L.~Popa, T.~Poutanen,
  G.~W. Pratt, G.~Pr\'{e}zeau, S.~Prunet, J.~L. Puget, J.~P. Rachen,
  C.~R\"{a}th, R.~Rebolo, M.~Reinecke, M.~Remazeilles, C.~Renault, A.~Renzi,
  S.~Ricciardi, T.~Riller, I.~Ristorcelli, G.~Rocha, C.~Rosset, A.~Rotti,
  G.~Roudier, J.~A. Rubi\~{n}o Mart\'{\i}n, B.~Rusholme, M.~Sandri, D.~Santos,
  G.~Savini, D.~Scott, M.~D. Seiffert, E.~P.~S. Shellard, T.~Souradeep, L.~D.
  Spencer, J.~L. Starck, V.~Stolyarov, R.~Stompor, R.~Sudiwala, F.~Sureau,
  P.~Sutter, D.~Sutton, A.~S. Suur-Uski, J.~F. Sygnet, J.~A. Tauber,
  D.~Tavagnacco, L.~Terenzi, L.~Toffolatti, M.~Tomasi, M.~Tristram, M.~Tucci,
  J.~Tuovinen, M.~T\"{u}rler, L.~Valenziano, J.~Valiviita, B.~Van~Tent,
  J.~Varis, P.~Vielva, F.~Villa, N.~Vittorio, L.~A. Wade, B.~D. Wandelt, I.~K.
  Wehus, M.~White, A.~Wilkinson, D.~Yvon, A.~Zacchei, and A.~Zonca, ``Planck
  2013 results. {XXIII}. isotropy and statistics of the {CMB}'', 2013.

\bibitem[Bojowald(1999)]{Bojowald1999Loop}
M.~Bojowald, ``Loop quantum cosmology i: Kinematics'',
  \href{http://xxx.lanl.gov/abs/gr-qc/9910103}{{\ttfamily gr-qc/9910103}}.

\bibitem[Atiyah and Sutcliffe(2001)]{Atiyah2001The-Geometry}
M.~Atiyah and P.~Sutcliffe, ``The geometry of point particles'', {\em
  Proc.Roy.Soc.Lond. A458 (2002) 1089-1116}, May 2001.

\bibitem[Atiyah and Sutcliffe(2003)]{Atiyah2003Polyhedra}
M.~Atiyah and P.~Sutcliffe, ``Polyhedra in physics, chemistry and geometry'',
  Mar 2003.

\bibitem[Gonz\'{a}lez et~al.(1992)Gonz\'{a}lez, Guinea, and
  Vozmediano]{Gonzalez1992The-Electronic}
J.~Gonz\'{a}lez, F.~Guinea, and M.~A.~H. Vozmediano, ``The electronic spectrum
  of fullerenes from the dirac equation'',
  \href{http://xxx.lanl.gov/abs/cond-mat/9208004}{{\ttfamily
  cond-mat/9208004}}.

\bibitem[Neto et~al.(2008)Neto, Guinea, Peres, Novoselov, and
  Geim]{Neto2008Electronic}
A.~H.~C. Neto, F.~Guinea, N.~M.~R. Peres, K.~S. Novoselov, and A.~K. Geim,
  ``The electronic properties of graphene'', {\em Reviews of Modern Physics}
  {\bfseries 81} Feb (2008) 109--162,
  \href{http://xxx.lanl.gov/abs/0709.1163}{{\ttfamily 0709.1163}}.

\bibitem[Wallace(1947)]{Wallace1947The-Band}
P.~R. Wallace, ``{The Band Theory of Graphite}'', {\em Physical Review Online
  Archive (Prola)} {\bfseries 71} May (1947) 622--634.

\bibitem[Jain(1989)]{Jain1989Compositefermion}
J.~K. Jain, ``Composite-fermion approach for the fractional quantum hall
  effect'', {\em Physical Review Letters} {\bfseries 63} Jul (1989) 199--202.

\bibitem[Chernodub(2010)]{Chernodub2010Superconductivity}
M.~N. Chernodub, ``Superconductivity of {QCD} vacuum in strong magnetic
  field'',  \href{http://xxx.lanl.gov/abs/1008.1055}{{\ttfamily 1008.1055}}.

\bibitem[Chernodub(2011{\natexlab{a}})]{Chernodub2011Can-nothing}
M.~N. Chernodub, ``Can nothing be a superconductor and a superfluid?'',
  \href{http://xxx.lanl.gov/abs/1104.4404}{{\ttfamily arXiv:1104.4404}}.

\bibitem[Chernodub(2011{\natexlab{b}})]{Chernodub2011Spontaneous}
M.~N. Chernodub, ``Spontaneous electromagnetic superconductivity of vacuum in
  strong magnetic field: evidence from the {Nambu\&\#45;\&\#45;Jona}-lasinio
  model'',  \href{http://xxx.lanl.gov/abs/1101.0117}{{\ttfamily arXiv:1101.0117}}.

\bibitem[Hossenfelder(2013{\natexlab{a}})]{Hossenfelder2013aPhenomenology}
S.~Hossenfelder, ``Phenomenology of space-time imperfection i: Nonlocal
  defects'', Sep 2013, \href{http://xxx.lanl.gov/abs/1309.0311}{{\ttfamily arXiv:1309.0311}}.

\bibitem[Hossenfelder(2013{\natexlab{b}})]{Hossenfelder2013bPhenomenology}
S.~Hossenfelder, ``Phenomenology of space-time imperfection ii: Local
  defects'', Sep 2013, \href{http://xxx.lanl.gov/abs/1309.0314}{{\ttfamily arXiv:1309.0314}}.

\end{thebibliography}\endgroup

\end{document}